\documentclass[prb,aps,twocolumn]{revtex4}
\usepackage{graphicx,color,latexsym}
\usepackage{dcolumn}
\usepackage{amsmath,amssymb,epsf,bm}
\begin{document}
\title{Spin pumping by a moving domain wall at the interface of an antiferromagnetic insulator and a two-dimensional metal}
\author{A.~G. Mal'shukov}
\affiliation{Institute of Spectroscopy, Russian Academy of Sciences, Troitsk, Moscow, 108840, Russia}
\begin{abstract}
A domain wall (DW) which moves parallel to a magnetically compensated interface between an  antiferromagnetic  insulator (AFMI) and a two-dimensional (2D) metal can pump spin polarization into the metal. It is assumed that localized spins of a collinear AFMI interact with itinerant electrons through their exchange interaction on the interface. We employed the formalism of Keldysh Green's functions for electrons which experience  potential and spin-orbit scattering on random impurities. This formalism allows a unified analysis of  spin pumping, spin diffusion and spin relaxation effects on a 2D electron gas. It is shown that the pumping of a nonstaggered magnetization into the metal film takes place in the second order with respect to the interface exchange interaction. At sufficiently weak spin relaxation this pumping effect can be much stronger than the first-order effect of the Pauli magnetism which is produced by the small nonstaggered exchange field of the DW.  It is shown that the pumped polarization is sensitive to the geometry of the electron's Fermi surface and increases when the wave vector of the staggered magnetization approaches the nesting vector of the Fermi surface. In a disordered diffusive electron gas the induced spin polarization follows the motion of the domain wall. It is distributed asymmetrically around the DW over a distance which can be much larger than the DW width.
\end{abstract}
\maketitle

\section{Introduction}

Antiferromagnets (AFM) have drawn growing interest recently due to their potential use for various spintronic applications. One of the most important characteristics of spintronic devices is their ability to transmit  and control spin polarization. From this point of view AFM materials demonstrate numerous interesting features. The progress made in this field was presented  in several reviews (see, for instance [\onlinecite{Baltz,Gomonay,Yan,Wadley}]). Considerable progress has been achieved in understanding of mechanisms for angular moment  transfer between spins of  localized and itinerant electrons in metallic AFM, as well as interface spin transfer between a normal metal and a metallic or insulating AFM [\onlinecite{Zelezny,Cheng,Saidaoui,Swaving,Takei,Nunez,Ohnuma}]. These mechanisms  allow to control localized spins of AFM, as well as spins of itinerant electrons. For instance, the  spin current of electrons produces the torque effect on the staggered AFM magnetization. Recent experimental studies have demonstrated that this torque results in rotation of the magnetization and its switching [\onlinecite{Zhang,Cogulu,Omari,Wadley}]. Alternatively, when the N\'{e}el order varies in time the magnetization can be pumped through the interface into the electron gas of a paramagnetic metal which makes a contact with an AFM [\onlinecite{Cheng,Frangou}]. In this case the spin polarization may be delivered to the interface by spin waves [\onlinecite{Cheng,Vaidya,Li,Wang}], or by moving topological defects, like DWs and skyrmions. Compared with ferromagnets, in AFM spin waves and topological spin textures exhibit much faster dynamics with lower energy dissipation. For instance, in antiferromagnetic insulators spin waves can propagate over  large (submicron) distances due to their relatively high lifetime [\onlinecite{Lebrun}], while  DWs can move much faster than in ferromagnets [\onlinecite{Gomonay2,Kim,Avci,Zhou,Velez}]. These outstanding features of conducting and insulating antiferromagnets form the basis for their future applications in spintronic devices.

So far, the activity in studying the spin pumping from an AFMI into a normal metal was focused on three dimensional (3D) metals. On the other hand, there is a great interest in  heterostructures which are combined of magnetic systems and 2D metals. In particular, this interest is caused by recent success in creating of various 2D van der Waals metallic and insulating systems. However, the problem of spin pumping from AFMI's space-time dependent spin textures into 2D metal films was not addressed in literature. At the same time, there are some significant distinctions between 3D and 2D cases. First of all, 2D electrons undergo scattering from an interfacial spin texture  which has the same 2D dimensionality. Therefore, constructive interference of spin dependent scattering amplitudes from two AFM sublattices can result in a strong enhancement of the scattering probability. This effect becomes important when the Fermi surface reveals nesting parts with roughly the same wave vector as that of the staggered magnetization. In contrast, in 3D systems such an interference effect is smeared out due to integration over $k_z$, where $k_z$ is the component of the electron's wave-vector which is perpendicular to the interface. One more specific feature of 2D systems is that electronic transport takes place along the interface, so that electrons are always in contact with localized spins of the AFM, while in 3D systems the angular moment, which electrons obtain from a dynamic AFMI texture, is carried away from the interface. Therefore, with a good accuracy a 3D metal can be considered as a spin sink for electrons, that is not true for 2D systems.  In the former case, the  angular moment transfer is controlled  by the so called interface spin mixing conductance [\onlinecite{Tserkovnyak,Cheng}] which is simply a local characteristic of a given interface. Such an approach can not be applied for the analysis of the spin transfer across the AFMI/2D metal interface, because it is closely related to the lateral transport of 2D electrons. Therefore, one needs a unified theory which combines quantum dynamics of 2D electrons with their interface scattering from a space-time dependent  spin texture of AFMI.

In order to reach this goal we employ the Keldysh [\onlinecite{Keldysh}] formalism of nonequilibrium Green's functions for a disordered 2D electron gas which interacts with localized spins of an adjacent AFMI by means of the exchange interaction $J$. This formalism is applied to the problem of the spin pumping by a domain wall which moves along the magnetically compensated surface of AFMI.  Besides the potential scattering from random impurities, the spin-orbit scattering of electrons will also be taken into account. The latter gives rise to relaxation of the spin polarization of itinerant electrons. As a result, in the diffusive regime the spin density distribution will diffusively evolve  in space and decrease in time with some spin relaxation rate.

This problem will be considered within a simple tight binding model where  2D and AFMI lattices form a commensurate contact.  The exchange interaction is treated within the perturbation theory which is valid as long as $J\ll E_F$, where $E_F$ is the Fermi energy of conduction electrons. The Fermi level is placed not too close to the van Hove singularity, where a gap in the electron band energy is formed due to the interface exchange interaction with AFMI [\onlinecite{Baltz}]. Within the perturbation theory the pumping of (nonstaggered) spin polarization into a normal metal by the staggered N\'{e}el magnetization takes place only in even orders of the perturbational expansion with respect to $J$. On the other hand, besides the staggered magnetization, a time dependent spin texture of a moving DW carries a small nonstaggered component [\onlinecite{Baryakhtar}] which is localized near the DW. In turn, due to the interface exchange interaction such a "ferromagnetic" magnetization polarizes spins of itinerant electrons in adjacent normal metal already in the first order of the expansion in $J$. However, as it will be shown, in a reasonable range of parameters the effect of second order perturbational terms may exceed considerably that of the first order ones, because these competing effects involve  very different physical mechanisms. Indeed, in the former case the angular moment, which is carried by a DW, is transferred through the interface to electrons. This process leads to accumulation of the electron's spin polarization near DW and, as it will be shown, the latter increases with the spin relaxation time.  In contrast, the first-order effect is simply the Pauli magnetization which does not depend so dramatically on the spin relaxation rate.

The article is organized in the following way. In Sec.II a general formalism  of the spin density response in a 2D disordered electron gas to a moving DW is expressed in terms of Green's functions, up to the second-order with respect to the exchange interaction. Sec. III is devoted to calculations of the spin polarization. The results are discussed in Sec. IV. Two section are added in the Appendix in order to clarify some details of calculations.

\section{General formalism}

\subsection{Basic equations}

In this section we express the spin polarization  in 2D electron gas as an  expansion over the exchange interaction between itinerant electrons and localized spins of an adjacent AFMI. By assuming that the 2D lattice of the normal metal is commensurate with the lattice of localized spins on the AFMI interface and that metal atoms make an on-top contact with atoms of the AFMI, the exchange interaction can be written in the form
\begin{equation}\label{M}
M=J\sum_i c^{+}_i\mathbf{S}_i(t)\cdot\bm{\sigma}c_i \,.
\end{equation}
where $c^{+}_i=(c^{+}_{i\uparrow},c^{+}_{i\downarrow})$ is the two-component creation operator of an electron whose spin projections are $\uparrow$, or $\downarrow$ and $c_i$ is the conjugate to $c^{+}_i$ destruction operator.  The vector $\mathbf{S}_i(t)$ represents a spin which is localized on the lattice site $\mathbf{r}_i$ and $\bm{\sigma}=(\sigma_x,\sigma_y,\sigma_z)$ is the vector of Pauli matrices. Since there are two sublattices, the lattice sites will be denoted as $i1$ and $i2$. Correspondingly, spins localized on these sublattices will be denoted as $\mathbf{S}_{i1}$ and $\mathbf{S}_{i2}$.  These spins are treated as classical variables satisfying the constraint $|\mathbf{S}_i(t)|=S$. In many practical situations $\mathbf{S}_i$ varies slowly  within each of two AFM sublattices. Therefore,  one may introduce two vector fields $\mathbf{m}_1(\mathbf{r},t)$ and $\mathbf{m}_2(\mathbf{r},t)$, which are defined on sublattices 1 and 2, respectively, where $\mathbf{m}_{1(2)}(\mathbf{r}_i,t)=\mathbf{S}_{i1(2)}/S$. The N\'{e}el order is given by the unit vector field  $\mathbf{n}(\mathbf{r},t)=(\mathbf{m}_1(\mathbf{r},t) - \mathbf{m}_2(\mathbf{r},t))/|\mathbf{m}_1(\mathbf{r},t) - \mathbf{m}_2(\mathbf{r},t)|$. Due to the strong exchange coupling of spins in different sublattices we have $\mathbf{m}_1(\mathbf{r},t) \simeq -\mathbf{m}_2(\mathbf{r},t)$. Therefore, the nonstaggered field $\mathbf{m}(\mathbf{r},t)=(\mathbf{m}_1(\mathbf{r},t) + \mathbf{m}_2(\mathbf{r},t))/2 \ll 1$. In this case by using the Landau-Lifshitz-Gilbert equation $\mathbf{m}$ can be expressed [\onlinecite{Baryakhtar}] in terms of $\mathbf{n}$, as
\begin{equation}\label{m}
\mathbf{m}(\mathbf{r},t))=\frac{1}{E_{\mathrm{ex}}}(\partial_t \mathbf{n}(\mathbf{r},t))\times\mathbf{ n}(\mathbf{r},t)))\,,
\end{equation}
where $E_{\mathrm{ex}}$ is the exchange energy of near-neighbor spins in AFM.
Further, by expressing the operators $c_i$ as $c_i=\sum_{\mathbf{k}}c_{\mathbf{k}}\exp i\mathbf{k}\mathbf{r}_i$, matrix elements of the exchange interaction in Eq.(\ref{M}) can be written in terms  of $\mathbf{n}_{\mathbf{f}}(t)$ and $\mathbf{m}_{\mathbf{f}}(t)$, which are spatial Fourier transforms of these fields with the wave vector $\mathbf{f}$. These matrix elements are given by
\begin{equation}\label{Mkk}
M_{\mathbf{k}+\mathbf{f},\mathbf{k}}(t)=JS[c^{+}_{\mathbf{k}+\mathbf{f}+\mathbf{G}}\mathbf{n}_{\mathbf{f}}(t)\bm{\sigma}c_{\mathbf{k}}+
c^{+}_{\mathbf{k}+\mathbf{f}}\mathbf{m}_{\mathbf{f}}(t)\bm{\sigma}c_{\mathbf{k}}] \,,
\end{equation}
where $\mathbf{G}=(G_x,G_y)$ is the umklapp vector which is associated with the AFM's staggered magnetization, so that for the square lattice $G_x=\pm \pi/a$ and $G_y=\pm \pi/a$. In contrast, the vector $\mathbf{f}$ is small, namely, $f \ll 1/a$, where $a$ is the lattice constant.  It is because relatively slow spatial variations of $\mathbf{n}(\mathbf{r},t)$ and $\mathbf{m}(\mathbf{r},t)$ are determined by the spin texture of DW whose width is assumed to be much larger than $a$.

\begin{figure}[tp]
\includegraphics[width=8 cm]{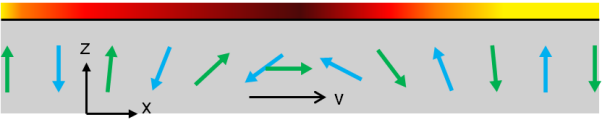}
\caption{A bilayer system is composed of an antiferromagnetic insulator and a 2D paramagnetic metal film. A domain wall which moves from left to right with the velocity $v$ leads to spin accumulation in the metal film. The spin is oriented parallel to the $y$-axis. The corresponding spin density (shown by color) is asymmetrically distributed around DW over distances which can be much larger than the width of the DW. } \label{fig1}
\end{figure}

The induced spin density of electrons  can be expressed in terms of the Keldysh [\onlinecite{Keldysh,Rammer}] function $G^K(\mathbf{r}.t;\mathbf{r}^{\prime},t^{\prime})$, which is a
2$\times$2 matrix in the spin space. This expression has the form
\begin{equation}\label{S}
\mathbf{S}(\mathbf{r},t)=-\frac{i}{4}\mathrm{Tr}[\bm{\sigma}\langle
G^K(\mathbf{r},t;\mathbf{r},t)\rangle_{\text{imp}}] \,,
\end{equation}
where $\langle...\rangle_{\text{imp}}$ denotes averaging over impurity positions. The Keldysh function, in turn, is given by the perturbational expansion over the exchange interaction $M$. The corresponding correction $\delta G^K(\mathbf{q},\omega)$ to the space-time Fourier transform of $G^K(\mathbf{r},t;\mathbf{r},t)$ can be written in terms of the unperturbed  Green's functions $G^K_{\mathbf{k},\mathbf{k}^{\prime}}(\epsilon)$, $G^r_{\mathbf{k},\mathbf{k}^{\prime}}(\epsilon)$ and $G^a_{\mathbf{k},\mathbf{k}^{\prime}}(\epsilon)$, which are, respectively, the Keldysh, retarded and advanced ones. These functions are not averaged over impurity positions. Therefore, they depend on the two wave vectors $\mathbf{k}$ and $\mathbf{k}^{\prime}$. They may be represented by the 2$\times$2 matrix $\hat{G}_{\mathbf{k},\mathbf{k}^{\prime}}(\epsilon)$ which is given by
\begin{equation}\label{G}
\hat{G}_{\mathbf{k},\mathbf{k}^{\prime}}(\epsilon)=\left[
\begin{matrix}
G^r_{\mathbf{k},\mathbf{k}^{\prime}}(\epsilon)&G^K_{\mathbf{k},\mathbf{k}^{\prime}}(\epsilon)\\
0 & G^a_{\mathbf{k},\mathbf{k}^{\prime}}(\epsilon)
\end{matrix}\right]\,,
\end{equation}
where in thermal equilibrium $G^K_{\mathbf{k},\mathbf{k}^{\prime}}(\epsilon)$ has the form
\begin{equation}\label{GK}
G^K_{\mathbf{k},\mathbf{k}^{\prime}}(\epsilon)=\left(G^r_{\mathbf{k},\mathbf{k}^{\prime}}(\epsilon)-G^a_{\mathbf{k},\mathbf{k}^{\prime}}(\epsilon)\right)\tanh(\epsilon/2k_BT)
\end{equation}
Within the Keldysh formalism [\onlinecite{Keldysh,Rammer}] the correction $\delta G^K(\mathbf{q},\omega)$ is given by
\begin{eqnarray}\label{deltaG}
&&\delta G^K(\mathbf{q},\omega)=\int \frac{d\epsilon}{2\pi}\sum_{\mathbf{k,p,p}^{\prime}}\left[\hat{G}_{\mathbf{k}^+,\mathbf{p}^+}(\epsilon^{+})\times \right.\nonumber \\ &&\left.
\hat{\Sigma}_{\mathbf{p},\mathbf{p}^{\prime}}(\epsilon,\omega,\mathbf{q})\hat{G}_{\mathbf{p}^{\prime -},\mathbf{k}^-}
(\epsilon^-)\right]^K\,,
\end{eqnarray}
where $\epsilon^{\pm}=\epsilon\pm\omega/2$, $\mathbf{k}^{\pm}=\mathbf{k}\pm\mathbf{q}/2$ and the superscript "K" denotes the Keldysh component of the matrix product in Eq.(\ref{deltaG}).  By performing the time Fourier transform of $\mathbf{n}_{\mathbf{f}}(t)$ and $\mathbf{m}_{\mathbf{f}}(t)$ in Eq.(\ref{Mkk}) one can express $\hat{\Sigma}$  in the form
\begin{eqnarray}\label{M1M2}
\hat{\Sigma}_{\mathbf{p},\mathbf{p}^{\prime}}(\epsilon,\omega,\mathbf{q})&=&\hat{\Sigma}^{(1)}(\omega,\mathbf{q})\delta_{\mathbf{p},\mathbf{p}^{\prime}}+\nonumber\\
&&\int \frac{d\nu}{2\pi}\sum_{\mathbf{f}}\hat{\Sigma}^{(2)}_{\mathbf{p},\mathbf{p}^{\prime}}(\epsilon,\omega,\mathbf{q};\nu,\mathbf{f})\,,
\end{eqnarray}
where the functions $\hat{\Sigma}^{(1)}$ and $\hat{\Sigma}^{(2)}$ correspond to the first-order and second-order corrections, respectively. They are given by
\begin{equation}\label{Sigma1}
\hat{\Sigma}^{(1)}(\omega,\mathbf{q})=\hat{1}JS
\bm{\sigma}\mathbf{m}_{\mathbf{q},\omega}
\end{equation}
and
\begin{eqnarray}\label{Sigma}
&&\hat{\Sigma}^{(2)}_{\mathbf{p},\mathbf{p}^{\prime}}(\epsilon,\omega,\mathbf{q};\nu,\mathbf{f})=2J^2S^2
\left(\bm{\sigma}\mathbf{n}_{\mathbf{f}^+,\nu^+}\right)\times\nonumber\\
&&\hat{G}_{\mathbf{p}-\mathbf{f}+\mathbf{G},\mathbf{p}^{\prime}-\mathbf{f}+\mathbf{G}}(\epsilon-\nu)
(\bm{\sigma}\mathbf{n}^*_{\mathbf{f}^-,\nu^-})\,,
\end{eqnarray}
where  $\nu^{\pm}=\nu\pm\omega/2$, $\mathbf{f}^{\pm}=\mathbf{f}\pm \mathbf{q}/2$ and $\hat{1}$ is the unit matrix in the Keldysh space. Here and below, $\mathbf{q}$ and $\omega$ are Fourier variables which are related to spatial and time variations of the induced spin density, while $\mathbf{f}$ and $\nu$ are associated with DW structure. They are "slow" variables. At the same time, $\mathbf{k},\mathbf{k}^{\prime}$ and $\mathbf{p},\mathbf{p}^{\prime}$ are relatively large electronic wave numbers, as well as $\epsilon$ is associated with the electron dynamics.

\begin{figure}[tp]
\includegraphics[width=5.5 cm]{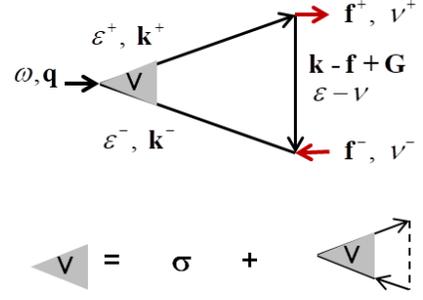}
\caption{a) A second-order Feynman diagram for the spin density $\mathbf{S}(\mathbf{q},\omega)$, which is induced by a moving domain wall. The DW perturbation Eq.(\ref{Mkk}) is shown by red arrows, $\mathbf{k}^{\pm}=\mathbf{k}\pm\mathbf{q}/2$, $\mathbf{f}^{\pm}=\mathbf{f}\pm\mathbf{q}/2$, $\epsilon^{\pm}=\epsilon\pm\omega/2$ and $\nu^{\prime \pm}=\nu^{\prime}\pm\omega/2$. b) The Bethe-Salpeter equation for the spin diffusion vertex $\mathbf{V}$, where $\bm{\sigma}$ is the vector of Pauli matrices. Elastic scattering from random impurirties is shown by the dashed line.} \label{fig2}
\end{figure}

The  second order contribution to $\mathbf{S}(\mathbf{r},t)$ is shown as a Feynman diagram in Fig.2, while the first-order term is given by a usual fermion loop.  The averaging in Eq.(\ref{S}) over random positions of  impurities results [\onlinecite{Altshuler}] in the occurrence of average Green functions and the vertex $\mathbf{V}$ in Fig.2. This  vertex describes multiple scattering processes which result in diffusion of particles and their spin relaxation. Within the Born approximation the calculation of $\mathbf{V}$ is reduced to a solving of the Bethe-Salpeter equation which is graphically shown in Fig.2. The multiple scattering processes are important when the frequency and momentum transfer  in the vertex are much smaller than  $\Gamma=1/2\tau$ and $1/l$, where $\tau$ and $l$ are the elastic scattering time and electron mean free path, respectively. Therefore, from Fig.2 it is seen that such a diffusion regime takes place when $\omega \ll \Gamma$ and $q\ll 1/l$. At the same time, the vertices which are associated with the exchange interaction of electrons with the staggered magnetization retain unrenormalized, because of the large momentum transfer $\sim \mathbf{G}$ caused by such a magnetization.
By using Eqs.(\ref{G},\ref{GK}) and Eqs.(\ref{Sigma1},\ref{Sigma}) one may express $\langle\delta G^K\rangle_{\mathrm{imp}}$ in Eq.(\ref{deltaG}), as well as the space-time Fourier transform $\mathbf{S}(\mathbf{q},\omega)$ of  Eq.(\ref{S}), in terms of averaged retarded and advanced Green's functions. The total spin polarization may be represented by a sum of terms which are renormalized by $\mathbf{V}$ and those which are not. The former will be denoted as $\mathbf{S}_V(\mathbf{q},\omega)$. It includes only  vertices which involve the product of retarded and advanced functions in the ladder series. Otherwise, the renormalization is not important [\onlinecite{Altshuler}]. At the same time the unrenormalized term is given by  the bare vertex $\bm{\sigma}$ instead of $\mathbf{V}$. The corresponding bare contribution to $\mathbf{S}(\mathbf{q},\omega)$ will be denoted as  $\mathbf{S}_0(\mathbf{q},\omega)$. By combining all terms, which are generated by the product of Keldysh matrices in  Eq.(\ref{deltaG}), we arrive at $\mathbf{S}=\mathbf{S}^{(1)}_V+\mathbf{S}^{(2)}_V+
\mathbf{S}^{(1)}_0+\mathbf{S}^{(2)}_0$. The spin densities, which enter into this sum, can be written in terms of the impurity averaged retarded and advanced Green's functions. Details are presented in Abstract B.

Eqs.(\ref{Sv1})-(\ref{Sigmaav}) form a basis for calculation of the spin density created by DW in 2D gas. The sum of the terms, that are given by Eqs.(\ref{Sv1}) and (\ref{S01}), represents the spin density induced by the interaction $JS\bm{\sigma}\mathbf{m}_{\mathbf{q},\omega}$ of electron spins with the nonstaggered Zeeman field $JS\mathbf{m}_{\mathbf{q},\omega}$. It is expressed in terms of the space-time dependent Pauli susceptibility, which is given by a single fermion loop, where the multiple scattering from impurities is taken into account in a standard way through the vertex function $\mathbf{V}$ [\onlinecite{Rammer,Altshuler}].  At the same time, Eqs.(\ref{Sv}) and (\ref{S0}) represent the effect whose nature is quite different from the Pauli magnetism. They describe second-order processes where the staggered magnetization, whose wave vector is $\mathbf{G}$, gives rise to quantum transitions of electrons between states with the wave vectors $\mathbf{k}$ and $\mathbf{k}+\mathbf{G}$. The DW, whose relatively smooth profile is characterized by the wave vector $\mathbf{f}$, such that $f \ll G$, adds $\mathbf{f}$ to $\mathbf{G}$. Therefore, the second-order scattering amplitude of electrons from a DW in an  antiferromagnet  carries terms of the form $(E_{\mathbf{k}+\mathbf{G}-\mathbf{f}}-E_{\mathbf{k}})^{-1}$, which are represented by $G^i_{\mathbf{k}-\mathbf{f}+\mathbf{G}}$ in Eq.(\ref{Sigmaav}). In this expression $\mathbf{k}$ is close to the Fermi surface. Therefore, $(E_{\mathbf{k}+\mathbf{G}-\mathbf{f}}-E_{\mathbf{k}})^{-1}$ becomes large if this surface is close to the nesting condition and $\mathbf{G}$ coincides with the nesting vector. This leads to the enhancement of the effect of second-order terms Eqs.(\ref{Sv}) and (\ref{S0}). Moreover, this expression becomes strongly dependent on $\mathbf{f}$, although the latter is much smaller than $\mathbf{G}$. On the other hand, when the Fermi surface is far from the nesting conditions this dependence is weak and the scattering amplitude may be expanded in powers of $\mathbf{v}_F\mathbf{f}$, where $\mathbf{v}_F$ is the Fermi velocity. Since $\mathbf{f}$ is associated with the coordinate dependence of the N\'{e}el order these terms generate spatial gradients of the form $\mathbf{n}(\mathbf{r},t)\times \nabla^i \mathbf{n}(\mathbf{r},t)$ in the induced spin density. Such sort of terms, with $\mathbf{n}(\mathbf{r},t)$ substituted for the ferromagnetic order parameter $\mathbf{m}(\mathbf{r},t)$, were discussed in connection with the spin pumping from a ferromagnetic DW into a 3D metal film [\onlinecite{Duine}].

\subsection{Disorder effects}

In this subsection we shall consider effects of disorder  on 2D electrons. Besides the usual potential scattering  from impurities  the spin-dependent scattering  will also be taken into account. The latter leads to spin relaxation of  electrons. A short-range  impurity scattering potential will be assumed. In this case, the scattering amplitude from a single impurity has the form [\onlinecite{Abrikosov}]
\begin{equation}\label{f}
f(\mathbf{k},\mathbf{k}^{\prime})=a+ib\bm{\sigma}\cdot(\mathbf{k}\times \mathbf{k}^{\prime})
\end{equation}
The first term in this expression is the isotropic spin-independent amplitude, while the second one represents the spin-orbit scattering. In a 2D system both incident and scattered wave vectors, $\mathbf{k}$ and $\mathbf{k}^{\prime}$, respectively, lie in the same $xy$ plane. Therefore, only the $\sigma_z$ Pauli matrix enters in the amplitude of the spin-orbit scattering. Therefore,  the scattering probability, which is presumably given by the second-order Born approximation over the scattering amplitude, is spin-independent. As follows from Ref.[\onlinecite{Abrikosov}], the total elastic scattering rate of electrons can be expressed as a sum of spin dependent and spin independent scattering channels. Accordingly, it can be written as
\begin{equation}\label{Gamma}
 \Gamma=1/2\tau=\pi N_F(a^2+ \frac{k^4_F}{2}b^2),
\end{equation}
where $\tau$ is the elastic scattering time, while $N_F$ and $k_F$ are  the state density and the Fermi wave-vector, respectively. At the same time, the retarded and advanced unperturbed Green's functions take the form [\onlinecite{Abrikosov}]
\begin{equation}\label{Gra}
G^r_{\mathbf{k}}(\epsilon)=G^a_{\mathbf{k}}(\epsilon)^*=\frac{1}{\epsilon-E_{\mathbf{k}}+\mu + i\Gamma}\,,
\end{equation}
where $\mu$ is the chemical potential.

Multiple scattering events should be taken into account in order to study the particle diffusion and spin relaxation effects. These effects are determined by the vertex function $\mathbf{V}(\omega,\mathbf{q})$ in Eqs.(\ref{Sv1}) and (\ref{Sv}). With spin-independent Green functions Eq.(\ref{Gra}) in hand the vertex $\mathbf{V}$ can be easy calculated by summation of ladder diagrams.  It is convenient to use its vector components $V^l$ ($l=x,y,z$) which are given by $V^l \mathbf{e}^l=(1/2)\mathrm{Tr}[\mathbf{V}\sigma^l]$ ($l=x,y,z$), where $\mathbf{e}^l$ is the unit vector in the $l$-direction. In more detail the calculation of  $V^l$ is presented in Appendix A.  Within the diffusion approximation, which is valid  at $\omega\ll \Gamma$  and $qv_F \ll \Gamma$, the sum of the ladder diagrams is given by the diffusion propagator
\begin{equation}\label{ladder}
V^l(\omega,\mathbf{q})=\frac{2\Gamma}{i\omega-Dq^2-\Gamma_s^l}\,,
\end{equation}
where the spin relaxation rates $\Gamma_s^l$ are  $\Gamma_s^x=\Gamma_s^y=\pi N_Fk^4_Fb^2$, $\Gamma_s^z=0$ and the diffusion constant $D=v_F^2\tau/2$.
It should be noted that the spin relaxation turns to zero when $l=z$ in Eq.(\ref{ladder}). It occurs because the spin projection on the $z$-axis is conserved due to a specific form of the spin-orbit scattering amplitude, which is proportional to $\sigma_z$ in a 2D gas. In this situation other mechanisms of the spin relaxation should be taken into account. However, such a strong spin relaxation channel as the scattering on AFM magnons can be efficient only at high enough temperatures. At low temperatures, due to the Fermi-liquid character of the electron gas, such an inelastic mechanism is weak, even in the absence of a gap in the excitation spectrum of magnons. The same can be said about the spin-lattice relaxation. The spin-orbit splitting of the conduction band might result in the spin relaxation through the D'yakonov-Perel mechanism [\onlinecite{DP}]. In the considered here simple model, however, such sort of the spin-orbit coupling does not take place. Nevertheless, a weak $\Gamma_s^z$ can be taken into account as a phenomenological parameter.

\section{Spin polarization of electrons}

\subsection{Spin pumping by the staggered magnetization}

Based on the general formalism presented in the previous section, let us consider the spin polarization which is produced  in the normal metal by the exchange field of AFMI in the presence of a moving DW. As it was discussed in Sec.II there are two contributions to the spin polarization. Namely, the first-order effect due to the nonstaggered "ferromagnetic" magnetization and the second-order one produced by the staggered exchange field. The latter effect, which is given by Eqs.(\ref{Sv}) and (\ref{S0}), will be considered in this subsection. Let us assume that within the classical theory the corresponding N\'{e}el vector $\mathbf{n}(\mathbf{r},t)$ is given by the well known solution of the equation of motion for a one-dimensional DW in an uniaxial AFM [\onlinecite{Walker}]. The precession of $\mathbf{n}(\mathbf{r},t)$ around the easy axis is assumed to be absent. It depends, however, on the method which is employed for the excitation of DW motion. For instance, the precession may be produced by the magnon's impact on the DW [\onlinecite{Tveten,Kim2}]. Otherwise, the azimuthal angle of the  N\'{e}el vector retains fixed during DW motion. The spin polarization of electrons, which can be induced by such DW, depends strongly on this angle. It becomes evident from the following consideration. Since $\hat{G}$ in Eq.(\ref{Sigma}) is given by the unit matrix in the spin space (retarded and advanced functions are given by the spin-independent $G^r$ and $G^a$ in Eq.(\ref{Gra})), the spin structure of the function $\hat{\Sigma}$ in Eq.(\ref{Sigma}) is determined by the product $(\bm{\sigma}\mathbf{n}_{\mathbf{f}^+,\nu^+})(\bm{\sigma}\mathbf{n}^*_{\mathbf{f}^-,\nu^-})$. Its spin-dependent part is given by $i\bm{\sigma}(\mathbf{n}_{\mathbf{f}^+,\nu^+}\times\mathbf{n}^*_{\mathbf{f}^-,\nu^-})$. Therefore, $\bm{\sigma}$ in this equation is perpendicular to the plane where the N\'{e}el vector of the DW resides. At the same time, this plane is fixed by the azimuthal angle of $\mathbf{n}$. In turn, as it follows  from Eqs.(\ref{Sv}),(\ref{P}) and  (\ref{ladder}) the trace in Eq.(\ref{Sv}) dictates that this spin direction must coincide with that of the vertex $\mathbf{V}$, whose vector components strongly depend on respective spin relaxation times. For example, if $\mathbf{n}(\mathbf{r},t)$ belongs to the $xy$ plane, the spin relaxation is given by $\Gamma^z_s$, which can be much smaller than $\Gamma^x_s$ and $\Gamma^y_s$ when the spin-orbit impurity scattering is a dominating mechanism of the spin relaxation.

Below, let us consider a different situation, such that the easy axis of AFM is oriented parallel to the $z$-direction, while the DW moves in the $x$-direction with the velocity $v$. In this case the vector $(\mathbf{n}_{\mathbf{f}^+,\nu^+}\times\mathbf{n}^*_{\mathbf{f}^-,\nu^-})$ lies in the $xy$-plane. Therefore, $\Gamma_s^x$ and $\Gamma_s^y$ enter in diffusion propagator Eq.(\ref{ladder}) . In the absence of precession around the easy axis the N\'{e}el vector is given in spherical coordinates by
\begin{equation}\label{neel}
\mathbf{n}(\mathbf{r},t)=(\sin\theta\cos\phi,\sin\theta\sin\phi,\cos\theta)\,,
\end{equation}
where $\cos\theta=\tanh[(x-vt)/\lambda]$ and $\lambda$ is the width of the DW, while $\phi$ is  fixed. Below, for simplicity  we choose $\phi=0$. In this case the functions $P^{ijk}$ in Eq.(\ref{P}) are proportional to the Pauli matrix $\sigma^y$. As a result, by taking the trace in Eqs.(\ref{Sv}) and (\ref{S0}) one obtains $\mathbf{Tr}[\mathbf{V}\sigma^y]=2V^y\mathbf{e}^y$. Therefore, spins  which are pumped into the 2D metal are oriented in the $y$-direction.

Parameters of the considered system are chosen in such a way that the wave vectors $q$ and $f$ in Eqs.(\ref{deltaG})-(\ref{Sigma}) are much smaller than $k_F$. Similarly, the frequencies $\omega$ and $\nu$ are much smaller than the Fermi energy. For a steady moving DW the variables $\omega$ and $q$, as well as $\nu$ and $f$, are related to each other. Indeed, the N\'{e}el vectors $\mathbf{n}_{\mathbf{f}^{\pm},\nu^{\pm}}$  in Eq.(\ref{Sigma}) can be written as
\begin{equation}\label{n}
\mathbf{n}_{\mathbf{f}^{\pm},\nu^{\pm}}=2\pi\delta(\nu^{\pm}-vf^{\pm}_x)\int d\xi e^{-if^{\pm}_x\xi}\mathbf{n}(\xi)\,,
 \end{equation}
where $\xi=x-vt$. The delta-function in this relation fixes the frequencies $\nu^{\pm}=vf^{\pm}_x$. By combining $\nu^{\pm}=\nu\pm \omega/2$ and $f_x^{\pm}=f_x\pm q_x/2$ we find that $\nu=vf_x$ and $\omega=vq_x$. Note, that $\omega$ and $q$ characterize the diffusion of electrons, because the diffusion propagator Eq.(\ref{ladder}) depends on these variables. At the same time,  $\nu$ and $f$ are associated with variations of  $\mathbf{n}(\mathbf{r},t)$ within a DW. The integration over wavevectors and frequencies in Eq.(\ref{Sv}) may be simplified at small $\omega$ and $\nu$ by replacing $\tanh(\Omega/k_BT)$ in Eq.(\ref{P}) with $\tanh(\Omega/k_BT)-\tanh(\epsilon/k_BT)$. Such a replacement does not change the result, because the terms in Eq.(\ref{Sv}) that are proportional to $\tanh(\epsilon/k_BT)$ cancel each other. Therefore, in any case, whether $\Omega=\epsilon\pm \omega/2$, or  $\Omega=\epsilon+ \nu$, at the low temperature such a replacement restricts the integration over $\epsilon$ to the range of small frequencies. As a result, the main contribution to the sum in Eq.(\ref{Sv}) is given by vectors $k$ which are close to the Fermi surface. Let us consider a simple tight binding model with the electronic band energy  $E_{\mathbf{k}}=-\alpha (\cos k_x a+\cos k_y a)$. Note, that in this case $E_{\mathbf{k}}=-E_{\mathbf{k}+\mathbf{G}}$. Consequently, by integrating Eq.(\ref{Sv}) over $\mathbf{k}$, in the leading approximation with respect to the small parameters $\omega/\Gamma,\nu/\Gamma,v_Fq/\Gamma,v_Ff/\Gamma$ and $\Gamma/\mu$,  we obtain the $y$-component $S^{y(2)}_V(\omega,\mathbf{q})$ of the vector  $\mathbf{S}^{(2)}_V(\omega,\mathbf{q})$ in the form
\begin{eqnarray}\label{Sv2}
S^{y(2)}_V(\omega,\mathbf{q})&=&-i\frac{J^2S^2}{\mu^2}V^y(\omega,\mathbf{q})N_F(\mu)\times\nonumber \\
&&\int \frac{d\nu}{2\pi}\sum_{\mathbf{f}}\nu(\mathbf{n}_{\mathbf{f}^+,\nu^+}\times\mathbf{n}^*_{\mathbf{f}^-,\nu^-})\,,
\end{eqnarray}
where $N_F(\mu)$ is electronic state density at the chemical potential. Details of this calculation can be found in Appendix B. It is seen that $S^{y(2)}_V$ gives the main contribution in the spin density $\mathbf{S}^{(2)}$ which is a sum of $S^{y(2)}_0(\omega,\mathbf{q})$  and $S^{y(2)}_V(\omega,\mathbf{q})$. Indeed, it follows from a comparison of Eqs.(\ref{S0}) and (\ref{Sv})  that these functions  differ from each other by the absence in $S^{y(2)}_0(\omega,\mathbf{q})$ of the diffusion propagator $V^y(\omega,\mathbf{q})$. The latter, however, is much larger than one, because in the diffusion regime $\Gamma$ is large in comparison with the denominator of Eq.(\ref{ladder}).  Therefore, $S^{y(2)}_0$ is small compared with $S^{y(2)}_V$, so that the total spin density $\mathbf{S}^{(2)}=\mathbf{S}^{(2)}_0+\mathbf{S}^{(2)}_V\simeq\mathbf{S}^{(2)}_V$. It is important to note that $\mathbf{S}^{(2)}$ is proportional to $\mu^{-2}$. Since the chemical potential is measured from the middle of the band, this dependence means that the pumping effect increases when $\mu$ approaches to the van Hove singularity. Such a dependence agrees with the discussed above role of the Fermi surface nesting in the spin pumping.

Note, that in the case of a 1D domain wall, which moves in the $x$-direction, the second line in Eq.(\ref{Sv2}) may be expressed as
\begin{eqnarray}\label{nn}
&&\int \frac{d\nu}{2\pi}\sum_{\mathbf{f}}\nu(\mathbf{n}_{\mathbf{f}^+,\nu^+}\times\mathbf{n}^*_{\mathbf{f}^-,\nu^-})=2\pi iv\times\nonumber \\
&&\delta(\omega-vq_x)\delta_{q_y}\int d\xi e^{-iq_x\xi}\left(\mathbf{n}(\xi)\times\nabla_{\xi}\mathbf{n}(\xi)\right)\,.
\end{eqnarray}
Therefore, by setting $\cos\theta=\tanh (\xi/\lambda)$ in Eq.(\ref{neel}) we obtain $(\mathbf{n}(\xi)\times\nabla_{\xi}\mathbf{n}(\xi))=(1/\lambda)\cosh^{-1}(\xi/\lambda)$. Further, from Eqs.(\ref{ladder}), (\ref{Sv2})  and (\ref{nn}) the spatial dependence of the induced spin density can be written as $S^{y(2)}(\mathbf{r},t)=S^{y(2)}(\xi)=S_V^{y(2)}(\xi)$, where
\begin{equation}\label{Sfin}
S^{y(2)}(\xi)=\frac{J^2S^2}{\mu^2}\int \frac{dqd\xi^{\prime}}{2\pi\lambda}\frac{N_Fv\Gamma}{\cosh\frac{\xi^{\prime}}{\lambda}}\frac{\exp iq(\xi-\xi^{\prime})}{Dq^2-ivq+\Gamma_s^y}\,.
\end{equation}
By calculating the integral over $q$ we arrive at
\begin{eqnarray}\label{Sfin2}
&&S^{y(2)}(\xi)=\frac{J^2S^2}{2\mu^2}\frac{\Gamma N_Fv}{\sqrt{v^2+4\Gamma_s^yD}}\int \frac{d\xi^{\prime}}{\lambda}\frac{1}{\cosh\frac{\xi^{\prime}}{\lambda}}\times\nonumber \\
&&\left(e^{-p_1|\xi-\xi^{\prime}|}\theta(\xi-\xi^{\prime})+e^{p_2|\xi-\xi^{\prime}|}\theta(\xi^{\prime}-\xi)\right)\,,
\end{eqnarray}
where $p_1=(1/2D)(v+\sqrt{v^2+4\Gamma_s^yD})$ and $p_2=(1/2D)(v-\sqrt{v^2+4\Gamma_s^yD})$. These parameters determine widths of forward ($1/p_1$) and backward ($1/p_2$) diffuse propagations of the spin density with respect to the DW center.  In most realistic cases these widths are much larger than the DW width $\lambda$. By taking into account that $D\sim v_F^2\tau$ it follows from above expressions for $p_1$ and $p_2$ that $p_1\lambda\ll 1$ and $p_2\lambda\ll 1$, if max$[v_F/v, 1/\sqrt{\Gamma_s\tau}]\gg \lambda/l$, where $l$ is the electron's elastic mean free path. Typically $v_F/v \gg 1$ and $\Gamma_s\tau \ll 1$. Therefore, $p_1\lambda$ and $p_2\lambda$ are small, if the ratio  $\lambda/l$ is not too large. In this case the integration over $\xi^{\prime}$ in Eq.(\ref{Sfin2}) is restricted to a  small interval around $\xi^{\prime}=0$. Hence, by setting $\xi^{\prime}=0$ in Eq.(\ref{Sfin2}) it may be simplified to
\begin{equation}\label{Sfin3}
S^{y(2)}(\xi)=\frac{J^2S^2}{2\mu^2}\frac{\pi\Gamma N_Fv}{\sqrt{v^2+4\Gamma_s^yD}}
\left(e^{-p_1\xi}\theta(\xi)+e^{p_2|\xi|}\theta(-\xi)\right)\,.
\end{equation}
Note, that  the singularity of the derivative of $S^y(\xi)$  at $\xi=0$  vanishes, when the actual profile of the DW in the integral over $\xi^{\prime}$ is taken into account. Such fine details, however, are not important, because the above expression is valid only in the range of distances from DW which are much larger than its width.

\subsection{Spin polarization due to the nonstaggered magnetization of the domain wall}

The polarization which is induced in the normal metal by the "ferromagnetic" part of the exchange field is given by Eqs.(\ref{Sv1}) and (\ref{S01}).  For a one-dimensional  DW moving  with the velocity $v$  in the $x$-direction  $\mathbf{m}(\mathbf{r},t)=\mathbf{m}(x-vt)\equiv \mathbf{m}(\xi)$. By performing Fourier transformation of $\mathbf{m}(\mathbf{r},t)$ in Eq.(\ref{m}) one can see that its Fourier transform $\mathbf{m}_{\mathbf{q},\omega}$ is given by the right-hand side of Eq.(\ref{nn}) multiplied by the factor $-i/E_{\mathrm{ex}}$. This expression should be substituted in Eq.(\ref{Sigma1}) which, in turn, enters in Eqs.(\ref{Sv1}) and (\ref{S01}) for the spin density. By taking into account Eq.(\ref{ladder}), in the leading approximation with respect to the small parameters $\omega/\Gamma$ and $v_Fq_x/\Gamma$ the spin density $S^{(1)}=S_V^{(1)}+S_0^{(1)}$ can be written in the form
\begin{eqnarray}\label{S(1)}
&&S^{y(1)}(\xi)=- i\frac{JS}{E_{\mathrm{ex}}}N_Fv\int \frac{dqd\xi^{\prime}}{2\pi}\frac{\exp iq(\xi-\xi^{\prime})}{\lambda\cosh\frac{\xi^{\prime}}{\lambda}}\times\nonumber\\
&&\left[\frac{vq}{Dq^2-ivq+\Gamma_s^y}-i\right]\,.
\end{eqnarray}
Similar to the previous subsection, at distances from DW which are larger than $\lambda$ the integration over $\xi^{\prime}$ may be simplified, that results in
\begin{eqnarray}\label{S(1)fin}
&&S^{y(1)}(\xi)= \frac{\pi JSN_Fv}{E_{\mathrm{ex}}}\times\nonumber\\
&&\left(\delta(\xi)+\frac{vp_1e^{-p_1\xi}\theta(\xi)+vp_2e^{p_2|\xi|}\theta(-\xi)}{\sqrt{v^2+4\Gamma_s^yD}}\right)\,.
\end{eqnarray}
The first term in this expression is represented by the delta-function, as long as large distances are of interest, while in the range of DW it is given by the function $\pi^{-1}\lambda^{-1}\cosh^{-1} (\xi/\lambda)$, as it follows from the second term in the square  brackets of Eq.(\ref{S(1)}).

\section{Discussion}

It was shown above that a moving DW in AFMI induces a macroscopic spin density in a 2D electron gas, which is in the epitaxial contact with the compensated surface of the AFMI. This spin density is a sum of two parts $\mathbf{S}^{(1)}$ and $\mathbf{S}^{(2)}$ whose $y$-components are given by Eqs.(\ref{Sfin2}-\ref{Sfin3}) and Eqs.(\ref{S(1)}-\ref{S(1)fin}). The spin polarization is perpendicular to the plane where the N\'{e}el vector of DW evolves. In the considered case it is the {zx}-plane, so that the polarization is parallel to the $y$-axis.  The spin densities $\mathbf{S}^{(1)}$ and $\mathbf{S}^{(2)}$ have very different physical origins. Thus, $\mathbf{S}^{(1)}$ is induced by the nonstaggered part of the exchange field of a moving DW. It is determined by the Pauli magnetism in the first order with respect to the exchange interaction. At the same time, $\mathbf{S}^{(2)}$ is produced by the staggered N\'{e}el order in the second order with respect to $J$. It is instructive to compare these competing  contributions to the total spin polarization $\mathbf{S}$. Let us consider the total polarization which is accumulated in the metal (per unit length of DW in the y-direction). By integrating spin densities Eq.(\ref{Sfin3}) and Eq.(\ref{S(1)fin}) over $x$ we obtain $S_{\mathrm{tot}}^{y(1)}=\pi (JS/E_{\mathrm{ex}})N_Fv$ and $S_{\mathrm{tot}}^{y(2)}=(J^2S^2/\mu^2) (\Gamma/\Gamma_s^y)N_Fv$. First, it should be noted a fundamental difference between these two spin densities. The effect of the staggered magnetization $S_{\mathrm{tot}}^{(2)}$ diverges when the spin relaxation rate $\Gamma_s \rightarrow 0$ [\onlinecite{feedback}], while $S_{\mathrm{tot}}^{(1)}$ does not depend on the spin relaxation. The reason is that the former effect is based on spin pumping from AFM into the metal. In the absence of spin relaxation such a pumping would lead to steady increase  of the spin polarization in the metal. However, it saturates due to the spin relaxation at some finite value. At the same time, the nonstaggered exchange field gives rise to the spin polarization through the Pauli mechanism. The motion of DW results only in redistribution of this polarization over the 2D metal. From a comparison of  $S_{\mathrm{tot}}^{y(1)}$ and $S_{\mathrm{tot}}^{y(2)}$ it is seen that the pumping mechanism dominates at $JS>(\Gamma_s^y/\Gamma)(\mu^2/E_{\mathrm{ex}})$. For instance, at $JS/\mu=0.1$ and $\mu \sim E_{\mathrm{ex}}$ the above inequality is fulfilled at $\Gamma_s^y/\Gamma \lesssim 0.1$, which always takes place if the spin relaxation is determined by the spin-orbit impurity scattering. This is because $\Gamma_s$ is a relativistic correction to $\Gamma$, which is given by the nonrelativistic potential scattering. Also, $\mu$ must be close enough to zero, where the van Hove singularity just in the middle of the band is placed. In the above evaluation $\mu$ was assumed to be comparable with the exchange energy of spins in AFM, but much larger than the exchange interaction between itinerant and AFM interface spins. Note, that the latter condition follows from the perturbation theory.  The unperturbed Green functions which were used above for the calculation of  Feynman diagrams do not take into account the interface exchange interaction $J$ of electrons with the unperturbed N\'{e}el order (without DW). On the other hand, this order leads to the gap in the middle of the band [\onlinecite{Baltz,Zelezny}] and modifies electronic wave functions. One may neglect these effects only for electronic states, which are sufficiently far from the gap. The electronic band with the gap is represented by two branches $\pm \sqrt{E^2_{\mathbf{k}}+S^2J^2}$. The effect of this gap on the Green function can be ignored if $S^2J^2\ll \mu^2$ and $E_{\mathbf{k}}\approx \mu$.  This is the main condition which restricts the strength of the second order effect.

As long as the effect of the staggered DW's magnetization  dominates, let us further focus  on the discussion of this effect. As it follows from Eqs.(\ref{Sfin2}) and (\ref{Sfin3}), the pumped spin density is distributed asymmetrically with respect to DW. A tail of spin polarized electrons extends behind the moving DW over the distance $\sim p_2^{-1}$, which increases up to $\infty$ when $v \gg v_F\sqrt{\Gamma_s\tau}$. At the same time, ahead of DW the spin density extends up to $\sim p_1^{-1}$ which is always smaller than $l(\sqrt{(v^2/v_F^2)+\Gamma_s\tau})^{-1/2}$. The magnitude of the induced magnetization is largest at the center of the DW. From Eq.(\ref{Sfin3}) one can evaluate it as
\begin{equation}\label{Smax}
S^y(0)= N_F\Gamma\frac{J^2S^2}{\mu^2}\frac{v}{\sqrt{v^2+4\Gamma_s^yD}}\,.
\end{equation}
This expression turns to 0 at $v=0$ and reaches its maximum when the DW velocity  $v\gg \sqrt{\Gamma_s^yD}\sim v_F\sqrt{\Gamma_s^y\tau}$, by taking into account that $D=v_F^2\tau/2$. It will be, however, assumed that $v$ stays less than the magnon's group velocity which plays the role of the light speed. So that relativistic effects, such as the Lorentz contraction of DW, can be ignored Refs.[\onlinecite{Kosevich,Kim2}]. Further, we notice that $S^y$ increases as $\mu^{-2}$ at $\mu \rightarrow 0$ where the Fermi level approaches the middle of the band. This position of $\mu$ corresponds to the nesting condition for the Fermi surface, when $E_{\mathbf{k}_F}=E_{\mathbf{k}_F+\mathbf{G}}$. It agrees with the qualitative behavior of the second-order scattering of 2D electrons from an AFMI spin texture, which was discussed in  SecIIA.  In order to evaluate quantitatively the effect of spin pumping by a DW, it is convenient to compare it with the spin polarization which could be produced in the electron gas by an external static magnetic field $H_{\mathrm{eff}}$. Since this spin polarization is $N_F\mu_B H_{\mathrm{eff}}$, we get $\mu_B H_{\mathrm{eff}}=S^y(0)/N_F$, where $S^y(0)$ is given by  Eq.(\ref{Smax}). From this equation one can see that  in the case when  $\Gamma_s^yD/v^2 \gg 1$ the effective magnetic field $\mu_BH_{\mathrm{eff}} \sim \Gamma (J^2S^2/\mu^2)(v/v_F)(\Gamma_s^y\tau)^{-1/2}$ (it was taken into account that $D=v^2_F\tau/2$). By assuming $\tau=10^{-13}$s, $\Gamma_s^y=10^{9}$s$^{-1}$, $v_F=10^5$m/s, $v=100$m/s, $(J^2S^2/\mu^2)$=0.01, and $\Gamma=(\hbar/2\tau)\sim 3$meV, we obtain $\mu_BH_{\mathrm{eff}}=0.003$meV, or $H_{\mathrm{eff}}\sim 0.05$T. In this parameter range the induced spin polarization  linearly increases with the velocity of DW. As shown in Refs.[\onlinecite{Tveten,Kim2}], in AFM insulators spin waves provide an efficient mechanism for DW propulsion.  A more pronounced effect may be realized due to the so called staggered torque effect, which is produced by electric current in a metallic AFM [\onlinecite{Gomonay2}]. Regarding the magnon's effect, the relatively high value $v\sim 100$m/s was calculated [\onlinecite{Tveten}] for circularly polarized magnons, while linearly polarized magnons produce much weaker effect. It should be noted that in the former case magnons cause precession of DW, so that the axial angle in Eq.(\ref{neel}) varies as $\phi=\Omega t$. As a result, the induced spin density will oscillate, in contrast to the stationary spin density soliton given by Eq.(\ref{Sfin3}). Moreover, the spin polarization vector will have not only $S^y$ components. Such a situation was not analyzed in this work. Also,  a further analysis is necessary of the physics close to the nesting point, as it was discussed in the end of Sec.IIA.

The spin polarization, which is pumped by DW, can be detected by measuring the electric current in a heavy metal contact. The contact may be placed at the right edge of the junction which is shown in Fig.1. The current can be produced by the inverse spin Hall effect, or spin galvanic effect due to the strong spin-orbit coupling of electrons in the heavy metal. This method is usually applied for detection of the pumped spin polarization Ref.[\onlinecite{Baltz}]. The other method, which we shall discuss in detail is based on the conversion of the spin current into electric one by passing through a ferromagnetic film Ref.[\onlinecite{Aronov,Johnson}]. In the considered set up a DW, which passes by the detector shown in Fig.3, injects the spin polarization into the contact and produces a pulse of the electric voltage there. The set up and corresponding calculations are presented in Appendix C. The voltage which is induced by the spin current is given by Eq.(\ref{phifin}). Let us evaluate it at $\Gamma$=30 meV, that corresponds to the elastic scattering time of 2D electrons $\tau\approx 1/2\Gamma=10^{-14}$s and the mean free path 10nm with the Fermi velocity 10$^6$m/s. This gives the diffusion constant $D_N\equiv D=v_F^2\tau/2=50$cm$^2$s$^{-1}$. Let us take $J^2S^2/\mu^2=0.1, N_F=m/2\pi\approx 2\times 10^{14}$eV$^{-1}$cm$^{-2}$, and $N_{FF}=2\times 10^{22}$eV$^{-1}$cm$^{-3}$. The latter is a typical DOS for 3d transition metals. At $d=10$nm the 2D DOS of the ferromagnetic film is $dN_{FF}=2\times10^{16}$eV$^{-1}$cm$^{-2}$. The DOS of the normal metal was taken according to the parabolic band model. Its enhancement with the approaching of the chemical potential to the van Hove singularity has been ignored. In the considered tight binding model this enhancement is large only when $\mu$ is very close to the antiferromagnetic gap (see e.g. Ref.[\onlinecite{Zelezny}]). Other parameters: $v$=100m/s, $\tilde{P}=0.1$, $\tau_{Ns}=\Gamma^{-1}_{Ns}\equiv 1/\Gamma_s^y=100$ps, $\tau_{Fs}=\Gamma^{-1}_{Fs}=10$ps, and $D_F=50$cm$^2$s$^{-1}$. With these parameters we obtain from Eq.(\ref{phifin}) $\Delta\phi \approx 30$nV. The parameters can vary significantly, so that $\Delta\phi$ can be smaller, or larger of the above evaluation. For the detection it is important to have a very thin ferromagnetic film, ideally a 2D film. Anyway, the above calculation shows that the spin pumping by a single DW can be detected.

This research is funded by the research project FFUU-2021-0003 of the Institute of Spectroscopy of the Russian Academy of sciences.

%%%%%%%%%%%%%%%%%%%%%%%%%%%%%%%%%%%%%%%%%%%%%%%%%%%%%%%%%%%%%%%
%%%%%%%%%%%%%%%%%%%%%%%%%%%%%%%%%%%%%%%%%%%%%%%%%%%%%%%%%%%%%%%

\appendix

\section{Calculation of the vertex function $\mathbf{V}(\mathbf{q},\omega)$}

A single element of the ladder array, which corresponds to the Bethe-Salpeter equation  in Fig. 2, is given by
\begin{eqnarray}\label{psi0}
&&\psi_0^{ij}=\frac{1}{2}\sum_{\mathbf{k}}\mathrm{Tr}[\sigma^i G^r(\epsilon^{+},\mathbf{k}^+)f(\mathbf{k}^+,\mathbf{k}^{+\prime})\sigma^j \times\nonumber\\
&&f(\mathbf{k}^{-\prime},\mathbf{k}^{-})G^a(\epsilon^{-},\mathbf{k}^-)]\,,
\end{eqnarray}
where $G^r$ and $G^a$ are given by Eq.(\ref{Gra}),  Since $\omega^{\prime} \ll \mu$ the integration in Eq.(\ref{psi0}) is restricted to $k\simeq k_F$. Therefore, one may set $|\mathbf{k}^{\pm}| =k_F$ and $|\mathbf{k}^{\pm\prime}| =k_F^{\prime}$ in the scattering amplitude $f$, which is given by  Eq.(\ref{f}). Note, that in general, for a nonspherical Fermi surface $k_F\neq k_F^{\prime}$. Further, as was noted in the main text, the spin-dependent part of $f$ is proportional to $\sigma_z$. Hence, by calculating the trace in Eq.(\ref{psi0}) one may express it in the form
\begin{eqnarray}\label{Tr}
\frac{1}{2}\mathrm{Tr}[\sigma^i f(\mathbf{k}^+,\mathbf{k}^{+\prime})\sigma^j
f(\mathbf{k}^{-\prime},\mathbf{k}^{-})]=a^2\delta^{ij}+&&\nonumber\\
b^2k_F^2k_F^{\prime 2}(1-(\mathbf{\hat{k}}\mathbf{\hat{k}}^{\prime})^2)(\delta^{iz}\delta^{jz}-\delta^{ix}\delta^{jx}-\delta^{iy}\delta^{jy})\,,&&
\end{eqnarray}
where $\mathbf{\hat{k}}$ and $\mathbf{\hat{k}}^{\prime}$ are unit vectors which are parallel to $\mathbf{k}$ and $\mathbf{k}^{\prime}$, respectively. Let us assume, for simplicity, that the Fermi line has an approximately circular form, which takes place if the chemical potential is sufficiently far from the middle of the considered tight binding band. In this case after integration in  Eq.(\ref{psi0}) over $|\mathbf{k}|$ and by taking into account Eq.(\ref{Tr}) we obtain
\begin{equation}\label{psi0finz}
\psi_0^{zz}=2\pi iN_F\int \frac{d\phi}{2\pi}\frac{a^2+b^2k_F^4(1-(\mathbf{\hat{k}}\mathbf{\hat{k}}^{\prime})^2)}{\omega-\mathbf{v}_F\mathbf{q}+2i\Gamma}
\end{equation}
and
\begin{equation}\label{psi0finx}
\psi_0^{xx}=\psi_0^{yy}=2\pi iN_F\int \frac{d\phi}{2\pi}\frac{a^2-b^2k_F^4(1-(\mathbf{\hat{k}}\mathbf{\hat{k}}^{\prime})^2)}{\omega-\mathbf{v}_F\mathbf{q}+2i\Gamma}\,,
\end{equation}
where $\phi$ is the polar angle of the vector $|\mathbf{\hat{k}}|$. These functions depend on the angle between $\mathbf{k}^{\prime}$ and $\mathbf{q}$. This dependence originates from the second term in integrands of Eqs.(\ref{psi0finz}) and (\ref{psi0finx}), which, in turn, is proportional to the spin-orbit scattering amplitude $b$. The latter is much weaker than the usual potential scattering $a$. Therefore, in the leading approximation the vertex function is angular independent. Hence, within this approximation  Eqs.(\ref{psi0finz}) and (\ref{psi0finx}) can be averaged over directions of $k^{\prime}$. By expanding them over $\omega$,  $q$ and $b$ up to their respective leading orders, after averaging over $\mathbf{\hat{k}}^{\prime}$ and by substituting $\Gamma$ from Eq.(\ref{Gamma}), we arrive at
\begin{eqnarray}\label{psi0fin2}
&&\psi_0^{xx}=\psi_0^{yy}=1+\frac{1}{2\Gamma}(i\omega-Dq^2-2\pi N_Fb^2k_F^4)\,;\nonumber\\
&&\psi_0^{zz}=1+\frac{1}{2\Gamma}(i\omega-Dq^2)\,.
\end{eqnarray}
By summing the ladder diagrams the vertex function $V^l$ can be expressed in terms of $\psi_0$ as
\begin{equation}\label{ladder2}
V^l=\frac{1}{1-\psi_0^{ll}}\,.
\end{equation}
Finally, by substituting $\psi_0^{ll}$ from Eq.(\ref{psi0fin2}) we obtain Eq.(\ref{ladder}).

\section{Calculation of the spin density}

As can be seen from Eqs.(\ref{deltaG}-\ref{Sigma}), $\delta G^K$ and, hence, the spin density are represented by products of retarded and advanced functions in various combinations. Thus, the product of two functions enter in $\mathbf{S}^{(1)}$, while that of three functions contributes in $\mathbf{S}^{(2)}$. The Keldysh component of a binary product has the form $[G_1G_2]^K=G_1^KG_2^a+G_1^rG_2^K$, while that of a triple product is $[G_1G_2G_3]^K=G_1^KG_2^aG_3^a+G_1^rG_2^KG_3^a+G_1^rG_2^rG_3^K$. The labels 1,2 and 3 denote variables of averaged Green functions. $G^K$ is given by  $G^K_{\mathbf{k}}(\epsilon)=[G^r_{\mathbf{k}}(\epsilon)-G^a_{\mathbf{k}}(\epsilon)]\tanh \epsilon/k_BT$. By following these rules we arrive at
\begin{eqnarray}\label{Sv1}
\mathbf{S}^{(1)}_V(\mathbf{q},\mathbf{\omega})=-\frac{i}{4}\sum_{\mathbf{k}}\int \frac{d\epsilon}{2\pi}\mathrm{Tr}[\mathbf{V}(\omega,\mathbf{q})G^r_{\mathbf{k}^+}(\epsilon^{+})\times &&\nonumber\\
\Sigma^{(1)}(\omega,\mathbf{q})G^a_{\mathbf{k}^-}(\epsilon^{-})]\left(\tanh\frac{\epsilon^{+}}{k_BT}-\tanh\frac{\epsilon^{-}}{k_BT}\right)\,,&&
\end{eqnarray}
\begin{eqnarray}\label{S01}
&&\mathbf{S}^{(1)}_0(\mathbf{q},\mathbf{\omega})=-\frac{i}{4}\sum_{\mathbf{k}}\int \frac{d\epsilon}{2\pi}\mathrm{Tr}\left[\bm{\sigma}G^r_{\mathbf{k}^+}(\epsilon^{+})\Sigma^{(1)}(\omega,\mathbf{q})\times\right.\nonumber\\
&&\left.G^r_{\mathbf{k}^-}(\epsilon^{-})\tanh\frac{\epsilon^{-}}{k_BT}-\bm{\sigma}G^a_{\mathbf{k}^+}(\epsilon^{+})\Sigma^{(1)}(\omega,\mathbf{q})\times\right.\nonumber\\
&&\left.G^a_{\mathbf{k}^-}(\epsilon^{-})\tanh\frac{\epsilon^{+}}{k_BT}\right]\,,
\end{eqnarray}
\begin{eqnarray}\label{Sv}
&&\mathbf{S}_V^{(2)}(\mathbf{q},\mathbf{\omega})=
-\frac{i}{4}\sum_{\mathbf{k},\mathbf{f}}\int \frac{d\epsilon}{2\pi}\frac{d\nu}{2\pi}\mathrm{Tr}\left[\mathbf{V}(\omega,\mathbf{q})(P^{raa}(\epsilon^{+}) \right.\nonumber \\
&&\left.-P^{rra}(\epsilon^{-})+ P^{rra}(\epsilon-\nu)-P^{raa}(\epsilon-\nu))\right]
\end{eqnarray}
and
\begin{eqnarray}\label{S0}
&&\mathbf{S}_0^{(2)}(\mathbf{q},\mathbf{\omega})=
-\frac{i}{4}\sum_{\mathbf{k},\mathbf{f}}\int \frac{d\epsilon}{2\pi}\frac{d\nu}{2\pi}\mathrm{Tr}\left[\bm{\sigma}(P^{rrr}(\epsilon^{-})- \right.\nonumber \\ &&\left.P^{aaa}(\epsilon^{+}))\right]\,,
\end{eqnarray}
where the functions $P^{ijk}(\Omega)$, with $\Omega=\epsilon^{\pm}$ and $\Omega=\epsilon -\nu$, are given by
\begin{eqnarray}\label{P}
&&P^{ijk}(\Omega)=G^i_{\mathbf{k}^+}\left(\epsilon^{+}\right)\Sigma^{(2)j}_{\mathbf{k}}(\epsilon,\omega,\mathbf{q};\nu,\mathbf{f})\times \nonumber \\ &&G^k_{\mathbf{k}^-}
(\epsilon^{-})\tanh\left(\frac{\Omega}{k_BT}\right)\,,
\end{eqnarray}
where $G^i_{\mathbf{k}}(\omega)=\langle G^i_{\mathbf{k},\mathbf{k}^{\prime}}(\omega)\rangle_{\mathrm{imp}}\delta_{\mathbf{k},\mathbf{k}^{\prime}}$, $\Sigma^{(2)i}_{\mathbf{k}}(\epsilon,\omega,\mathbf{q};\nu,f)=\langle\Sigma^{(2)i}_{\mathbf{k},\mathbf{k}^{\prime}}(\epsilon,\omega,\mathbf{q};\nu,f)\rangle_{\mathrm{imp}}
\delta_{\mathbf{k},\mathbf{k}^{\prime}}$ and $i=r,a$. It follows from Eq.(\ref{Sigma}) that after the averaging over impurity positions the  function $\Sigma^{(2)i}_{\mathbf{k},\mathbf{k}^{\prime}}(\epsilon,\omega,\mathbf{q};\nu,f)$ becomes
\begin{eqnarray}\label{Sigmaav}
&&\Sigma^{(2)i}_{\mathbf{k}}(\epsilon,\omega,\mathbf{q};\nu,f)=2J^2S^2
\left(\bm{\sigma}\mathbf{n}_{\mathbf{f}^+,\nu^+}\right)\times\nonumber\\
&&G^i_{\mathbf{k}-\mathbf{f}+\mathbf{G}}(\epsilon-\nu)
(\bm{\sigma}\mathbf{n}^*_{\mathbf{f}^-,\nu^-})\,.
\end{eqnarray}

Further, we calculate the spin density, starting from Eq.(\ref{Sv}). There are four terms, namely $P^{raa}(\epsilon^{+}),P^{rra}(\epsilon^{-}), P^{rra}(\epsilon+\nu)$, and  $P^{raa}(\epsilon+\nu)$ in the right hand side of this equation. Each of these terms is proportional to the Fermi statistical factors in the form of $\tanh[(\epsilon+\omega_i)/2k_BT]-\tanh[\epsilon/2k_BT]$, where $\omega_i=\pm \omega$, or  $\omega_i= \nu$. Since these frequencies are small, all these terms  give small contribution to the spin density. However, the first two of them are much smaller than the third one. The reason is that the former are proportional to $\omega$, while the latter $\sim \nu$. However, the function  $\Sigma^i_{\mathbf{k}}(\epsilon,\omega;\nu,\mathbf{f})$ in Eq.(\ref{P}) is an odd function with respect to change of signs of $\nu$ and $\mathbf{f}$, because two N\'{e}el vectors in Eq.(\ref{Sigma}) form a cross product, as it was explained in subsection III A. Indeed, $(\mathbf{n}_{\mathbf{f}^+,\nu^+}\times\mathbf{n}^*_{\mathbf{f}^-,\nu^-})$ changes sign when $\nu\rightarrow -\nu, \mathbf{f}\rightarrow -\mathbf{f}$, because at such sign reversal $\mathbf{f}^{\pm}\rightarrow -\mathbf{f}^{\mp}$ and $\nu^{\pm}\rightarrow -\nu^{\mp}$. Hence, $\mathbf{n}_{\mathbf{f}^+,\nu^+}\leftrightarrows \mathbf{n}^*_{\mathbf{f}^-,\nu^-}$. Therefore, those terms which are proportional to $\omega$ have an additional small factor, because they must turn to zero with $\nu$ and $f$. In contrast, the term $\sim \nu$ is initially an odd function of $\nu$. On this reason, it dominates in Eq.(\ref{Sv}), as long as $\nu \Gamma \gg \omega\cdot \mathrm{max}(\nu,v_Ff)$. By using Eq.(\ref{Sigma}) the contribution of this term in Eq.(\ref{Sv}) can be written in the form
\begin{eqnarray}\label{Sv3}
&&\mathbf{S}^{(2)}_V(\mathbf{q},\mathbf{\omega})=J^2S^2
\frac{1}{2}\sum_{\mathbf{f}}\int \frac{d\nu}{2\pi}I(\omega,\mathbf{q};\nu,\mathbf{f})\times\nonumber \\ &&\mathrm{Tr}\left[\mathbf{V}(\omega,\mathbf{q})\left(\bm{\sigma}\cdot(\mathbf{n}_{\mathbf{f}^+,\nu^+}\times\mathbf{n}^*_{\mathbf{f}^-,\nu^-})\right)\right]\,,
\end{eqnarray}
where
\begin{widetext}
\begin{eqnarray}\label{I}
I(\omega,\mathbf{q};\nu,\mathbf{f})=\sum_{\mathbf{k}}\int \frac{d\epsilon}{2\pi}G^r_{\mathbf{k}^+}(\epsilon^+)G^a_{\mathbf{k}^-}
(\epsilon^{-})
\left(G^r_{\mathbf{k}+\mathbf{f}+\mathbf{G}}(\epsilon+\nu)-G^a_{\mathbf{k}+\mathbf{f}+\mathbf{G}}(\epsilon+\nu)\right)\left(\tanh\frac{\epsilon+\nu}{k_BT}-
\tanh\frac{\epsilon}{k_BT}\right)\,.
\end{eqnarray}
\end{widetext}
By substituting in this equation the expressions for Green functions from Eq.(\ref{G}) and by expanding  there the electron energy as $E_{\mathbf{k}+\mathbf{Q}}=E_{\mathbf{k}}+\mathbf{v}_{k}\cdot \mathbf{Q}$, where $\mathbf{Q}=\mathbf{q}$, or $\mathbf{f}$ and $\mathbf{v}_{\mathbf{Q}}$ is the velocity, the integral $I$ can be written as
\begin{widetext}
\begin{eqnarray}\label{I2}
I(\omega,\mathbf{q};\nu,\mathbf{f})=\int \frac{d^2k}{4\pi^2}\int \frac{d\epsilon}{2\pi}\frac{1}{(\epsilon+\frac{\omega}{2}-E_{\mathbf{k}}-\frac{\mathbf{v}_{\mathbf{k}}\mathbf{q}}{2}+\mu+i\Gamma)}
\frac{1}{(\epsilon-\frac{\omega}{2}-E_{\mathbf{k}}+\frac{\mathbf{v}_{\mathbf{k}}\mathbf{q}}{2}+\mu-i\Gamma)}\times&&\nonumber\\
\left(\frac{1}{(\epsilon+\frac{\nu}{2}-E_{\mathbf{k+G}}-\frac{\mathbf{v}_{\mathbf{k+G}}\mathbf{f}}{2}+\mu+i\Gamma)}-
\frac{1}{(\epsilon+\frac{\nu}{2}-E_{\mathbf{k+G}}-\frac{\mathbf{v}_{\mathbf{k+G}}\mathbf{f}}{2}+\mu-i\Gamma)}\right)\left(\tanh\frac{\epsilon+\nu}{k_BT}-
\tanh\frac{\epsilon}{k_BT}\right)\,.&&
\end{eqnarray}
\end{widetext}
One should take into account that within the tight binding model $E_{\mathbf{k+G}}=-E_{\mathbf{k}}$ and $v_{\mathbf{k+G}}=-v_{\mathbf{k}}$. Then, it is seen that since $|\epsilon| \sim |\nu|\ll \mu$, as well as $\omega$, $\Gamma$, $v_{\mathbf{k}}q$ and $v_{\mathbf{k+G}}f \ll \mu$ the major contribution to the integral at small $T$ is given by two poles, whose wave vectors are close to the Fermi surface. Therefore, one can set $v_{\mathbf{k}} \simeq v_{\mathbf{k}_F}\equiv v_F$, where $E_{\mathbf{k}_F}=\mu$. In the leading approximation $\mu \gg \Gamma \gg [v_Fq,v_Ff,\nu, \omega]$ this integration gives
\begin{equation}\label{Ifin}
I(\omega,\mathbf{q};\nu,\mathbf{f})=-i\frac{\nu}{\mu^2}N_F\,.
\end{equation}
By substituting this expression in Eq.(\ref{Sv3}) we arrive at Eq.(\ref{Sv2}).
\begin{figure}[bp]
\includegraphics[width=7 cm]{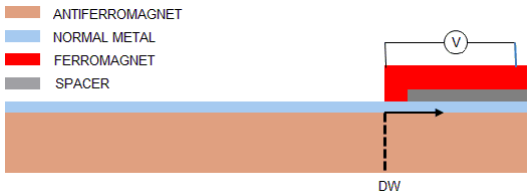}
\caption{A proposed set up for the detection of the spin polarization induced by a domain wall. The instant position of the DW is shown by the dashed line. The arrow shows the DW motion direction. The arrows origin is fixed at $x=0$.} \label{fig2}
\end{figure}
\section{Detection of the pumped spin}

The setup is shown in Fig.3. If spin-flip effects are weak, the current in the ferromagnet $J_F$ can be decomposed in the spin-up $J_{F\uparrow}$ and spin-down currents. Each of them is given by
\begin{eqnarray}\label{Jupdown}
J_{F\uparrow}&=&-D_{F\uparrow}e\nabla_x c_{F\uparrow}-\sigma_{F\uparrow}\nabla_x\phi \nonumber \\
J_{F\downarrow}&=&-D_{F\downarrow}e\nabla_xc_{F\downarrow}-\sigma_{F\downarrow}\nabla_x\phi\,,
\end{eqnarray}
 where $D_{F\uparrow}$ and $D_{F\downarrow}$ are diffusion constants in each of the spin subbands and  $\sigma_{F\uparrow}$, $\sigma_{F\downarrow}$ are conductivities. These currents are driven by gradients of the electron densities $c_{F\uparrow}$ and $c_{F\downarrow}$, as well as by the electric potential $\phi$. Since in general conductivities and diffusion constants  of electrons with opposite spin projections are different in ferromagnets, spin and charge transport characteristics become interdependent. Therefore, the spin polarized current can induce the electric current and potential, and vice versa [\onlinecite{Aronov,Johnson}]. This phenomenon serves as the basis for detecting spin currents. Since in the considered example the spin density, which is produced by DW, is directed parallel to the $y$-axis, we consider the ferromagnet whose magnetization is parallel to this axis. Therefore, $\rho_{Fs}$ and $J_{Fs}$ correspond to spins oriented along the $y$-axis.

 Let us denote the total electric current and the spin current as $J_F=J_{F\uparrow}+J_{F\downarrow}$ and $J_{Fs}=J_{F\uparrow}-J_{F\downarrow}$, respectively. Similarly, for charge and spin densities we have $\rho_F=ec_{F\uparrow}+ec_{F\downarrow}$ and $\rho_{Fs}=ec_{F\uparrow}-ec_{F\downarrow}$. By combining two equations in (\ref{Jupdown}) and taking into account the charge neutrality constrain $\rho_F=0$ we arrive at
\begin{eqnarray}\label{transport}
J_{F}&=&-D_{Fs}\nabla_x \rho_{Fs}-\sigma_{F}\nabla_x\phi \nonumber \\
J_{Fs}&=&-D_{F}\nabla_x\rho_{Fs}-\sigma_{Fs}\nabla_x\phi \,,
\end{eqnarray}
where $D_{F}=(D_{F\uparrow}+D_{F\downarrow})/2$, $D_{Fs}=(D_{F\uparrow}-D_{F\downarrow})/2$, $\sigma_{F}=\sigma_{F\uparrow}+\sigma_{F\downarrow}$ and $\sigma_{Fs}=\sigma_{F\uparrow}-\sigma_{F\downarrow}$. Since in the open circuit the electric current $J_F=0$, Eq.(\ref{transport}) gives
\begin{equation}\label{phi}
\sigma_{F}\nabla_x\phi=-D_{Fs}\nabla_x \rho_{Fs}\,.
\end{equation}
By substituting this equation in the second line of Eq.(\ref{transport}) we obtain
\begin{equation}\label{Dtilda}
J_{Fs}=-\tilde{D}_{F}\nabla_x\rho_{Fs}\,,
\end{equation}
where $\tilde{D}_{F}=D_F(1-PD_{Fs}/D_F)$, with $P$ denoting the polarization coefficient $P=(\sigma_{F\uparrow}-\sigma_{F\downarrow})/(\sigma_{F\uparrow}+\sigma_{F\downarrow})$.
In the following, this unimportant renormalization of the diffusion coefficient will be ignored and we set $\tilde{D}_{F}=D_{F}$.  The spin diffusion equation has the conventional form:
\begin{equation}\label{diff}
\frac{\partial\rho_{Fs}}{\partial t}=-\nabla_x J_{Fs}-\Gamma_{Fs}\rho_{Fs}\,,
\end{equation}
where $\Gamma_{Fs}$ is the spin relaxation rate. By taking into account Eq.(\ref{Dtilda}) and performing the time Fourier transform this equation can be written in the form
\begin{equation}\label{diff2}
-i\omega\rho_{Fs}=D_F\nabla^2_x \rho_{Fs}-\Gamma_{Fs}\rho_{Fs}\,.
\end{equation}
Further, Eq.(\ref{phi}) can be integrated over the ferromagnetic region  $x>0$. This integration results in an equation which expresses the electric voltage via the spin density on the border between the ferromagnet and 2D metal. This relation has the form
\begin{equation}\label{phi2}
\sigma_{F}\Delta\phi=D_{Fs}\rho_{Fs}(0)\,.
\end{equation}
where $\Delta\phi$ is the voltage which is measured by the voltmeter shown at Fig.3. It was taken into account that $\rho(x) \rightarrow 0$ when $x\gg\sqrt{D_{F}/\Gamma_{Fs}} $. If the magnetic film is thin, so that its thickness $d$ is much less than the spin diffusion length ($d\ll \sqrt{D_{F}/\Gamma_{Fs}}$), the spin density is homogeneously distributed over $d$. Therefore, $d\rho_{Fs}$ can be considered as a 2D spin density. The same can be said about the current $dJ_{Fs}$. The spin density should be calculated from Eq.(\ref{diff2}) with the boundary conditions
\begin{equation}\label{bc}
d\rho_{Fs}(0)=\rho_{Ns}(0)\,\,, dJ_{Fs}(0)=J_{Ns}(0)\,,
\end{equation}
 where  $J_{Ns}(0)$ and $\rho_{Ns}(0)$ are the spin current and spin density on the normal metal side ($x<0$) of the boundary. The above boundary conditions are valid if a significant interface spin relaxation is absent.  In the ferromagnet the relation between $J_{Fs}(0)$ and $\rho_{Fs}(0)$ can be  simply found from the solution of Eq.(\ref{diff2}) which has the form
 \begin{equation}\label{rhox}
 \rho_{Fs}(x)=\rho_{Fs}(0)e^{-x/l_{Fs}(\omega)},
\end{equation}
 where $l^{-1}_{Fs}=\sqrt{(\Gamma_{Fs}-i\omega) D_F^{-1}}$. Further, Eq.(\ref{Dtilda}) gives the sought relation
\begin{equation}\label{Jrho}
J_{Fs}(0)=\rho_{Fs}(0)D_F/l_{Fs}
\end{equation}

The spin diffusion equation in the normal metal can be written as
\begin{equation}\label{diff3}
-i\omega\rho_{Ns}=D_N\nabla^2_x \rho_{Ns}-\Gamma_{Ns}\rho_{Fs} + W(\omega,x)\,.
\end{equation}
This equation follows from Eqs.(\ref{Sv2}) and (\ref{nn}) by converting them into the coordinate representation. In Eq.(\ref{diff3}) the parameters $\rho_{Ns}$, $\Gamma_{Ns}$ and $D_N$ are identified with, respectively,  $S^y$, $\Gamma^y_s$  and $D$ in the main text. The so calculated source term $W$ has the form
\begin{equation}\label{W}
W(\omega,x)=N_F\Gamma v \frac{J^2S^2}{\mu^2}\int dt\frac{1}{\lambda\cosh\frac{x-vt}{\lambda}}e^{i\omega t}\,.
\end{equation}
As it will be clear below, for chosen values of $v$ and $\Gamma_{Ns}$ the characteristic length $v/\omega$ is of the order of 10$^2$ nm, that is much larger than the typical DW width $\lambda$. Therefore, one can set $t=\omega/v$ in the exponential function in Eq.(\ref{W}). So, we finally obtain
\begin{equation}\label{W2}
W(\omega,x)=\pi N_F\Gamma \frac{J^2S^2}{\mu^2}e^{i\omega x/v}\,.
\end{equation}

The general solution of  Eq.(\ref{diff3})  at $x<0$ is a sum of  two functions. One of them is produced by the source and the other is originated from the interface with the ferromagnet. The former ($\rho^{(1)}_{Ns}$) is given by the time Fourier transform of Eq.(\ref{Sfin3}), while the latter ($\rho^{(2)}_{Ns}$) is obtained from  Eq.(\ref{diff3}) in the form
\begin{equation}\label{rho2}
\rho^{(2)}_{Ns}(x)=Ae^{-|x|/l_{Ns}}\,,
\end{equation}
where $l^{-1}_{Ns}=\sqrt{(\Gamma_{Ns}-i\omega) D_N^{-1}}$. It was assumed that width of the contact between the ferromagnet and 2D film is much smaller than $l^{-1}_{Ns}$ which is in a submicron range. Therefore, it is possible to treat the contact as a point at $x=0$. As a consequence, $\rho^{(2)}$ takes a simple form (\ref{rho2}).
The coefficients $A$ and $\rho_{Fs}(0)$ in Eq.(\ref{rhox}) and Eq.(\ref{rho2}) can be found from the boundary conditions Eq.(\ref{bc}) and Eq.(\ref{Jrho}). In the considered set up we have for the current
\begin{equation}\label{bc2}
J_{Ns}(0^-)-J_{Ns}(0^+)=dJ_{Fs}(0)\,.
\end{equation}
From this equation, by taking into account Eqs.(\ref{rho2}) and (\ref{Jrho}) we obtain
\begin{equation}\label{rhofs}
d\rho_{Fs}(0)=\frac{v^2W(\omega,0)}{D_N(vp_1+i\omega)(vp_2+i\omega)}\frac{1}{1+(l_{Ns}/2l_{Fs})}\,,
\end{equation}
where the factors $p_1$ and $p_2$ are defined just below Eq.(\ref{Sfin2}). Further, the electric potential which is induced by the spin current in the ferromagnetic film can be  calculated from  Eq.(\ref{phi}). The time dependence of the measured voltage has a form of a pulse whose shape is determined by the spatial dependence of  the spin density soliton, which is represented by Eq.(\ref{Sfin3}). In the center of this pulse at $t=0$ the spin density at $x=0$ can be calculated by integrating  Eq.(\ref{rhofs}) over $\omega$. Since $p_1>0$ and $p_2<0$, $\rho_{Fs}(0)$ has a single pole in the upper complex semiplane, while another pole and cuts are in the lower semiplane. Since the integral is converging, by calculating the residue at $\omega=ivp_1$ and by substituting the result in Eq.(\ref{phi}) we finally obtain
\begin{equation}\label{phifin}
e\Delta\phi|_{t=0}=r\Gamma \tilde{P}\frac{\pi N_F }{2N_{FF}d}\frac{J^2S^2}{\mu^2}\frac{v}{\sqrt{v^2+4\Gamma_{Ns}D_N}}\,,
\end{equation}
where $N_{FF}=(N_{F\uparrow}+N_{F\downarrow})/2$ is the density of electron states at the Fermi level per one spin in the ferromagnet. The dimensionless factor $r$ is given by
\begin{equation}\label{r}
r=\left[\frac{1}{1+(l_{Ns}/2l_{Fs})}\right]_{\omega=ivp_1}
\end{equation}
and
\begin{equation}\label{Ptilda}
\tilde{P}=\frac{D_{F\uparrow}-D_{F\downarrow}}{D_{F\uparrow}+D_{F\downarrow}}
\end{equation}
is the polarization factor for the diffusion coefficient of the magnet. When calculating Eq.(\ref{phifin}) the Einstein relation $\sigma_F=2e^2D_FN_{FF}$ was employed. The potential $\Delta\phi|_{t=0}$ is evaluated for a set of material parameters in the end of Sec.IV.

%%%%%%%%%%%%%%%%%%%%%%%%%%%%%%%%%%%%%%%%%%%%%%%%%%%%%%%%%%

%%%%%%%%%%%%%%%%%%%%%%%%%%%%%%%%%%%%%%%%%%%%%%%%%%%%%%%%%%%%%%%%%%%%%%%%%%%%%%%%%%%%%%%

\begin{thebibliography}{99}
\bibitem{feedback}
When the pumped polarization becomes too large the feedback effect should be taken into account. This situation needs an additional analysis.
\end{thebibliography}
\end{document}